\documentclass[prl,twocolumn,aps,epsf]{revtex4}
\usepackage{epsfig}

\newcommand{\ep}{\epsilon}
\newcommand{\sig}{\sigma}

\newcommand{\tta}{\theta}

\newcommand{\BDG}{Bogoliubov-de Gennes}

\begin{document}

\title{Incomplete Andreev reflection in a clean SFS junction}

\author{J\'{e}r\^{o}me Cayssol $^{1,2}$ and Gilles Montambaux $^{1}$}

\affiliation{(1)Laboratoire de Physique des Solides, Associ\'e au CNRS, Universit\'e
Paris Sud,  91405 Orsay, France}

\affiliation{(2)Laboratoire de Physique Th\'eorique et Mod\`eles Statistiques, Associ\'e au CNRS, Universit\'e
Paris Sud,  91405 Orsay, France}

\begin{abstract}
We study the Josephson effect in a clean
Superconductor-Ferromagnet-Superconductor junction for arbitrarily large spin polarizations. The Andreev
reflection at a clean Ferromagnet-Superconductor interface is incomplete, and Andreev channels with a 
large incidence angle are progressively
suppressed with increasing exchange energy. As a result, the critical current exhibits oscillations as a function of the exchange energy and of the length of the ferromagnet and has a temperature dependence which deviates from the one predicted by the quasiclassical theory. 
\end{abstract}
\maketitle


Current understanding of the Superconductor-Ferromagnet-Superconductor (SFS) Josephson effect is limited to small spin polarizations. In the case of conventional 
superconductors, the Josephson current is due to the
Andreev \cite{andreev64} conversion of singlet Cooper pairs into
correlated electrons and holes with opposite spins propagating coherently in the
ferromagnetic metal. Applying the Eilenberger equations \cite{eilen68} to a clean 
multichannel SFS junction, Buzdin {\it et al.} \cite{buzdin82} have predicted that this non dissipative current 
oscillates as a function of both the exchange energy splitting $E_{ex}$ and the length
$d$ of the ferromagnet, because of the mismatch $2 E_{ex}/\hbar v_F$ between the spin-up
and spin-down Fermi wavevectors. This quasiclassical result assumes that the Andreev reflection is complete, 
as it is fully justified for weakly spin-polarized ferromagnetic alloys $E_{ex} \ll E_F$, $E_F$ being the Fermi energy. First experimental evidence for such oscillating 
critical current has recently been reported in Nb-Cu-Ni-Cu-Nb junctions \cite{blum02}. The so-called $\pi$-phase state of a SFS junction \cite{bulaev77} has also 
been observed using diffusive weak ferromagnetic alloys such as Cu$_{1-x}$Ni$_x$ \cite{ryazanov01} or Pd$_{1-x}$Ni$_x$ \cite{kontos01},\cite{guichard03},\cite{bauer03}.

In the new field of spintronics, devices with high spin polarization are used in order to manipulate spin polarized currents. In the recently discovered half metals (HM), such as CrO$_2$ and La$_{0.7}$Sr$_{0.3}$MnO$_3$, the current is completely spin polarized because one spin subband is insulating. Ferromagnetic elements Fe, Co, Ni, also exhibit quite large spin polarizations. Anticipating the interest for large spin polarizations, de Jong and Beenakker \cite{jong95} have shown that in this case the Andreev reflection is not complete at a clean Ferromagnet-Superconductor (FS) interface, in contrast to the case of a clean nonmagnetic Normal metal-Superconductor (NS) interface. {\it Even in the absence of impurity scattering}, normal reflection may occur because of the diagonal exchange potential barrier between the ferromagnet and the superconductor. This suppression of the Andreev reflection affects preferentially the channels with large transverse momentum. As a result, the sub-gap conductance of a ballistic FS
contact decreases quasi-linearly as a function of the spin polarization $\eta=E_{ex}/E_F$ from twice the normal state conductance ($\eta=0$) to zero ($\eta=1$), because of
the progressive suppression of the Andreev process. Using this principle, a point-contact Andreev reflection technique 
has been developed in order to mesure directly the spin polarization of materials \cite{soulen98},\cite{upadhyay98}, such as La$_{0.7}$Sr$_{0.3}$MnO$_3$,\, CrO$_2$,\, NiFe,\, NiMnSb, which were 
not easily accessible by spin resolved tunneling spectroscopy \cite{meservey73}. A huge amount of theoretical efforts has been devoted to transport properties in a nanoscale 
FS contact \cite{zutic99},\cite{zutic00},\cite{mazin99},\cite{kopu04} while few studies have considered the thermodynamical properties of FS heterostructures \cite{halterman01},\cite{eschrig03}.    

In this Letter, we address the physics of the Josephson effect {\it in a clean multichannel SFS junction in the range of 
arbitrarily large spin polarization}. We show how the Josephson current is modified by the ordinary reflection induced by the ferromagnet in the crossover from a SNS ($\eta=0$) to a S/HM/S junction ($\eta >1$). With increasing exchange
energy, the Andreev reflection is
suppressed for electrons propagating with a large incidence,
so that the number of channels contributing to the
total current decreases. This reduction of the number of "Andreev active
channels" has furthermore a subtle effect on the Josephson current: although the FS conductance is always reduced when $\eta$ increases \cite{jong95}, the critical current has a non-monotonic behavior, depending on 
the current-phase relationship of the
suppressed channels. For large spin polarizations, the 
oscillations of the critical current depend {\it separately} on the product $k_F d$ and on the spin polarization $\eta$. They are {\it reduced and shifted} with respect 
to the predictions of the quasiclassical theory \cite{buzdin82} in which only a single parameter, $2 E_{ex}d/(\hbar v_F) =\eta k_F d $, is relevant. For small spin polarizations, we naturally recover the quasiclassical results. In the HM limit $E_{ex} \rightarrow E_F$, the critical current vanishes because the Andreev reflection is totally suppressed for all the transverse channels. In addition, we study the temperature dependence of the critical current for different values of the spin polarization and of the length $d$ of the ferromagnet.


We consider a clean short SFS junction with a large number $M$ of transverse channels and 
with a length $d$ of the ferromagnetic region much smaller than the coherence length of the superconductor $\xi_o =
\hbar v_F /\Delta_o$, where $\Delta_o$ is the $T=0$ superconducting gap. The itinerant ferromagnetism is described
within the Stoner model by an effective one body potential $V_\sig (x)=-\sig
E_{ex}$ which depends on the spin direction, characterized by $\sig =\pm
1$. In the superconducting leads,
$V_\sig (x)=0$. The superconducting pair potential is
$\Delta(x)=\mid \Delta \mid e^{i \chi/2}$ in the left lead 
and $\Delta(x)=\mid \Delta \mid e^{-i \chi/2}$ in the right lead. In the
absence of spin-flip scattering, the \BDG\, equations split in two sets of independent equations for 
the spin channels ($u_\uparrow ,v_\downarrow$) and ($u_\downarrow , v _\uparrow$) 
\begin{equation}\label{eqbogosfs}
\left(
\begin{array}{cc}
H_o + V_\sig (x)    &  \Delta(x) \\ \Delta(x)^* & -H_{o}^{*}+ V_\sig (x)      \\
\end{array}
\right) \left(
\begin{array}{l}
 u_\sig \\ v_{-\sig}
\end{array}
\right) =\ep(\chi) \left(
\begin{array}{l}
 u_\sig \\ v_{-\sig}
\end{array}
\right),
\end{equation}
where $\ep(\chi)$ is the quasiparticle energy mesured from the
Fermi energy \cite{Gennes}. The kinetic part of the  
Hamiltonian $H_o = [\left(-i \hbar d/dx-qA(x)\right)^{2} -E_F
]/2m $, with the
effective mass of electron and hole $m$, is expressed in terms of 
the vector potential $A(x)$, which is responsible for the phase difference $\chi$ between the
leads, and $E_F =\hbar^2 k_{F}^{2} /2m$ is the Fermi energy. The Fermi velocities are identical in 
both superconductors and in the paramagnetic metal. 

Because both the pair and the disorder potential are identically
zero in the ferromagnet, the eigenvectors of Eq. (\ref{eqbogosfs})
are electrons and holes with plane wave spatial dependencies. For a given 
transverse channel, the electron and hole longitudinal wavevectors, $k_{n \sig}$ and 
$h_{n -\sig}$ respectively, satisfy

\begin{eqnarray}
  \frac{\hbar^2 k_{n \sig}^2 }{2m}+E_n &=& E_F+\ep+\sig E_{ex} , \nonumber \\
  \frac{\hbar^2 h_{n -\sig}^2 }{2m}+E_n &=& E_F-\ep-\sig E_{ex} ,
  \label{vondepre1}
\end{eqnarray}
where $E_n$ is the transverse energy of the channel. One
may label the transverse channels by an angle $\tta_n$ which is
the incidence angle of the corresponding quasiparticle trajectory
\begin{equation}
E_n = \frac{\hbar^2 k_F^2}{2m} \sin^2 \tta_n =E_F
\sin^2 \theta_n.
\end{equation}

From Eq. (\ref{vondepre1}), one sees that an electron 
with incidence $\tta_n$ cannot form an Andreev bound state with a hole if 
$E_n =E_F \sin^2 \tta_n > E_F - E_{ex}$. Therefore the electron 
is normally reflected as an electron with the same spin for 
angle $\tta_n > \tta_\eta = \arccos \sqrt{\eta}$. Such a process is insensitive to the superconducting 
phase and thus carries no Josephson current. In the opposite case $\tta_n << \tta_\eta $, the 
Andreev reflection is complete and supports a finite current. In the following, the former kind of channel is referred to as "Andreev inactive" and the latter as "Andreev active".

Recently, we have performed detailed studies of the spectrum of a single
channel SFS junction for arbitrarily large exchange energies \cite{cayssol04}. Solving the \BDG\, equations, the spectrum is found to be strongly modified in comparison to the quasiclassical spectrum \cite{kuple90} 
because {\it gaps open at
$\chi=0$ and $\chi=\pi$}. However, due to a cancellation between the corrections associated 
to each anticrossing, the current {\it is almost unaffected} up to very large spin polarizations $\eta \approx 0.95$. The region in which Andreev reflection and
ordinary reflection coexist is extremely small. As a result, the Josephson current through a single channel SFS junction is given to great accuracy by the formula for perfect Andreev reflection \cite{buzdin82}
 \begin{eqnarray}\label{currentprb}
i(\chi,k_F d, \eta ,\tta_n=0)&=&\frac{\pi \Delta}{\phi_o} \sum_{\sig=\pm 1} \sin \frac{\chi+\sig a}{2}\\
&&\tanh \left( \frac{\Delta}{2T} \cos \left(\frac{\chi+\sig a}{2}\right)\right)\nonumber ,
\end{eqnarray}
for $\eta <1$ and it is zero for $\eta>1$. The parameter $a=(\sqrt{1+\eta}-\sqrt{1-\eta}) k_F d$\, is the phase shift accumulated between an electron and a hole located at the Fermi level during their propagation on a length $d$. 

In the present paper, we generalize this result to transverse channels with finite angle $\tta_n$, in the more realistic case of a finite width SFS junction. The crossover between Andreev active and 
inactive channels occurs in a narrow window of incidences in the vicinity of $\tta_\eta = \arccos \sqrt{\eta}$.  Below this
cut-off, the current carried by a single 
Andreev active channel is
\begin{eqnarray}\label{currentone}
i(\chi,k_F d, \eta ,\tta_n)&=&\frac{\pi \Delta}{\phi_o} \sum_{\sig=\pm 1} \sin \frac{\chi+\sig a_n}{2}\\
&&\tanh \left( \frac{\Delta}{2T} \cos \left(\frac{\chi+\sig a_n}{2}\right)\right)\nonumber ,
\end{eqnarray}
and it is zero for $\tta_\eta > \arccos \sqrt{\eta}$. In order to treat large exchange splitting, one 
has to take into account the exact 
band structure (here a simple isotropic parabolic band) and to express the phase shift between an electron and its Andreev reflected hole by
\begin{equation}\label{shift}
a_n =k_F d \cos \tta_n \left( \sqrt{ 1 + \frac{\eta}{\cos^2 \tta_n} }-\sqrt{ 1 - \frac{\eta}{\cos^2 \tta_n} } \right),
\end{equation}
instead of using the linearized form
\begin{equation}\label{shiftlin}
a_{n}=\frac{\eta k_F d}{ \cos \tta_n}=\frac{2 E_{ex} d}{\hbar v_F \cos \tta_n} .
\end{equation}
\begin{figure}[ht!]
\begin{center}
\epsfxsize 5.5cm \hspace*{-0.7cm} \epsffile{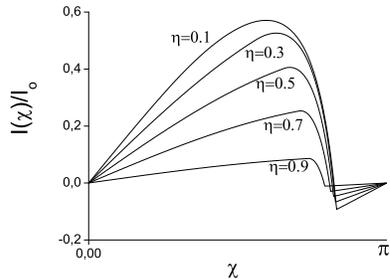}
\end{center}
\vspace*{-0.7cm}
\caption{{\it Current-phase relationships at zero temperature for
$a=\pi/4$ obtained for several pairs $(\eta,k_F d)$. In the quasiclassical approximation, the 
current is a function of the single parameter $a$ and does not decrease with increasing $\eta$. The current is given in
units of $I_o = \pi \Delta_0 /(e R_N)$ }}\label{fig1}
\end{figure}
The transverse channels considered above are independent because $V_\sig (x)$ 
is translationaly invariant in the transverse directions. Thus, the 
total current is the sum of the currents carried by each of them. As we assume a large
number of channels, the discrete sum over $n$ can be
replaced by an integral over the angle $\tta$. Calculating the
total current, one has to restrict the integration over Andreev active
levels only, so that the angular integral has to be
limited by the upper cut-off $\tta_\eta =\arccos \sqrt{\eta}$
\begin{equation}\label{currentsum}
I(\chi,k_F d,\eta)=\frac{k_F^2 S}{2 \pi} \int_{0}^{\tta_\eta} d\tta
\sin \tta \cos \tta \,\,i(\chi,k_F d,\eta,\tta) ,
\end{equation}
where $S$ is the
cross section area of the ferromagnet.

This expression, together with Eqs. (\ref{currentone}, \ref{shift}), is 
the central result of this Letter. It gives the Josephson current $I(\chi,k_F d,\eta)$ 
of a clean SFS junction in {\it the regime of arbitrarily large spin polarization}. Examples of current-phase relationships are shown in Fig. \ref{fig1}. In the limit of small polarization $\eta=E_{ex}/E_F \rightarrow 0$, we recover 
the quasiclassical current-phase relationship \cite{buzdin82} in which all the transverse channels contribute because 
$\tta_\eta \rightarrow \pi/2$.
\begin{figure}[ht!]
\begin{center}
\epsfxsize 5.5cm \hspace*{-0cm} \epsffile{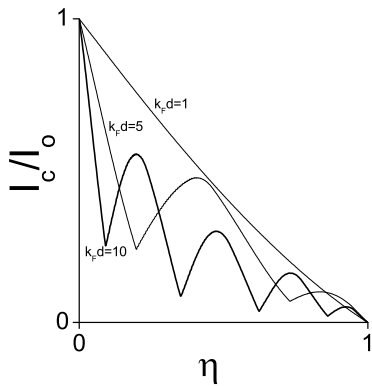}
\end{center}
\vspace*{-0.8cm}
\caption{{\it Zero temperature critical current $I_c (\eta)$ as
a function of $\eta=E_{ex}/E_F$ for different lengths of the ferromagnet,
$k_F d =1,\, 5, \, 10$. The current is given in
units of $I_o = \pi \Delta_0 /(e R_N)$.}}\label{fig2}
\end{figure}

\begin{figure}[ht!]
\begin{center}
 \epsfxsize 7.0cm \epsffile{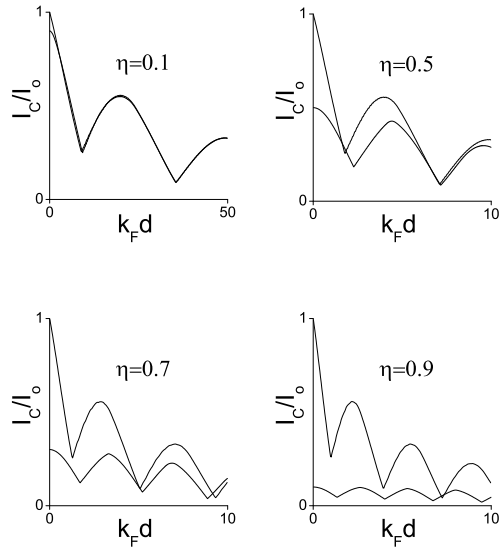}
\end{center}
\vspace*{-0.8cm}
\caption{{\it Zero temperature critical current $I_c (\eta)$ as
a function of $k_F d$ (thick lines), for different values of the spin polarization $\eta$. As $\eta$ increases, the
exact current deviates from the quasiclassical estimate (dashed lines). The current is given in
units of $I_o = \pi \Delta/(e R_N)$.}}\label{fig3}
\end{figure} 
Increasing the spin polarization, we study how the critical current evolves from the case of a weakly spin polarized junction to the S/HM/S junction. As shown in Fig. \ref{fig2}, the critical current has a non trivial oscillatory
behavior as a function of exchange splitting for a given length, namely for fixed $k_F d$. The number of oscillations occuring during the crossover from the SNS ($\eta=0$) to the S/HM/S junction ($\eta=1$) decreases 
when $k_F d$ is lowered. In the limit of an ultra-small junction $k_F d \approx 1$, there are no oscillations because the phase shift in Eq. (\ref{shift}) tends to zero, and all transverse channels carry the 
same SNS current with maximal value $i_o = 2 \pi \Delta /\phi_0$, where $\phi_o =h/e$ is the flux quantum. Consequently, the reduction 
of the total current is only governed by the upper cut-off in Eq. (\ref{currentsum})
\begin{equation}
I_c= M i_0 (1 - \eta)= {\pi \Delta \over e R_N}(1 -\eta).
\label{currentmax}
\end{equation} 
This linear reduction of the
current with increasing the exchange field is quite reminiscent of
the almost linear reduction obtained in Ref. \cite{jong95} for the
conductance of a FS nanocontact \cite{2d}. The total number of
transport channels $M=k_F^2 S/4 \pi$ is large and determines the
small normal state resistance $R_N =h/(2 e^2 M)$ of the
heterojunction. The natural unit for the critical current is $I_o=\pi \Delta/ e R_N $, namely 
the one of a short clean SNS junction.

Fig. \ref{fig3} represents the critical current as a function of the length $d$ of 
the ferromagnetic region, for different spin polarizations. We find that the oscillations are reduced and shifted with respect to the quasiclassical calculation. There are two reasons for these deviations. Firstly, trajectories with large incidence are progressively suppressed. Secondly, the 
phase shift between electrons and holes for a given channel [Eq. (\ref{shift})] depends on the particular band structure and differs from the linearized version $a_n=\eta k_F d/\cos \tta_n$. For large $d$, the oscillations decay  slowly at zero temperature. 
In real situations, they are expected to be severely reduced when $d$ exceeds the thermal length $L_T = \hbar v_F /T$ or the phase coherence length $L_\phi (T)$.

\begin{figure}[ht!]
\begin{center}
  \epsfxsize 7.0cm \epsffile{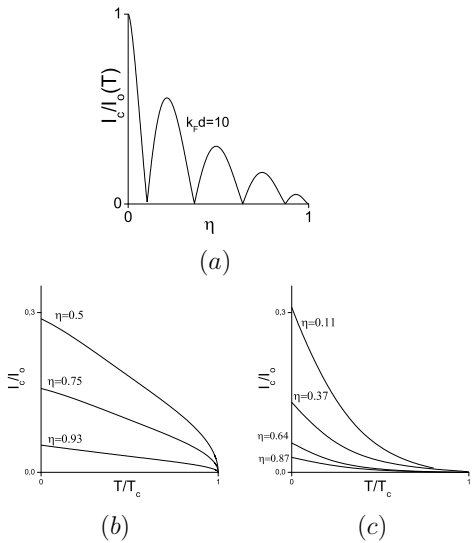}
\end{center}
\vspace*{-0.8cm}
\caption{{\it a) Critical current as a function of the spin polarization $\eta$ at $T=0.9 \, T_c$. It vanishes for particular values of the spin polarization, when the junction undergoes a $0-\pi$ transition. $I_0 (T) = \pi \Delta(T)^2 /(4 e R_N T_c)$ is the critical current for a SNS junction. b) Critical current [in units of $I_0 =\pi \Delta_o /(e R_N)$] as a function of the reduced temperature
$T/T_c$ for values of\, $\eta$ corresponding to the maxima of Fig (a). c) Critical current as a function of $T/T_c$ for different values of $\eta$  corresponding to the $0-\pi$ transitions.   All curves correspond to a short junction with $k_F d=10$.}}\label{fig4}
\end{figure}
We finally consider the effect of a finite temperature on the critical current. We have adopted the BCS temperature dependence of the
order parameter $\Delta(T)=\Delta_o \tanh (1.74 \sqrt{T_c /T -1})$, and the exchange energy 
is assumed to be temperature independent. For $T\approx T_c$, Fig. \ref{fig4}(a) shows that the critical current  oscillates with the spin polarization $\eta$ and cancels out for some values of $\eta$. In this temperature range, the current-phase relationship is sinusoidal $I(\chi)=I_c \sin \chi$ and the current vanishes identically when $I_c$ is zero. These cancellations are associated to transitions between the $0$-phase state and the $\pi$-state of the junction. For fixed parameters $k_F d$ and $\eta$, the critical current
decreases monotonously with increasing temperature $T$, as shown in Figs.
\ref{fig4}(b,c). This temperature dependence is very sensitive to the spin polarization. For polarizations corresponding to $0-\pi$ transitions, $I_c(T)$ decreases exponentially with temperature [Fig. \ref{fig4}(c)] whereas a much more slower decrease is obtained 
for the local maxima of the critical current [Fig. \ref{fig4}(b)].  

We have studied the Josephson current of a clean SFS junction 
for arbitrary large spin polarizations. The two physical effects involved are the reduction
of the number of active levels participating in the Andreev
process and the use of the non-linearized band structure. In any experiment with strong ferromagnetic elements or nearly half metallic compounds, the critical current oscillations should be affected by these effects. Firstly, the oscillations depend separately on the spin polarization $\eta$ and on the product $k_F d$ instead of the single combination $\eta k_F d$ as suggested by the quasiclassical theory. Secondly, when the temperature is increased from zero to the critical temperature, the 
local minima of the current are more strongly suppressed than the local maxima.

We thank Igor Zutic for useful comments.


\begin{thebibliography}{99}
\bibitem{2d} For a two-dimensional junction and in the 
limit $k_F d \ll 1$, the reduction of
the critical current with the exchange field is given by 
$I_c= M i_0 \sqrt{1 - \eta}= {\pi \Delta \over e R_N}\sqrt{1 -\eta}$, where $M$ is the number of transverse channels in a two-dimensional stripe.
\end{thebibliography}

\end{document}